\documentclass[referee]{svjour3}
\usepackage{epsfig}
\usepackage{graphicx}
\usepackage{lineno}
\usepackage{amsmath}
\usepackage{url}
\usepackage{amssymb}
\usepackage{subeqnarray,amsmath,amsfonts,graphicx}
\journalname{submitted to: International Journal of Dynamics and Control}

\def\dfrac{\displaystyle\frac}

\def\2q{\quad\quad}
\def\3q{\quad\quad\quad}

\def\<{\left\langle}
\def\>{\right\rangle}

\def\:{{\,:\,}}

\def\tot1{\sigma_{ij}^{(1,T)}}
\def\tot2{\sigma_{ij}^{2,T}}
\def\tot{\sigma_{ij}}

\begin{document}

\title{A numerical simulation of the COVID-19 epidemic  in Argentina using the SEIR model}

\titlerunning{COVID-19 in Argentina}

\author{Juan E. Santos  \and Jos\'{e} M. Carcione  \and Gabriela B. Savioli  \and Patricia M. Gauzellino 
\and Alejandro Ravecca \and Alfredo Moras
}

\institute{J. E. Santos \at 
         School of Earth Science and Technology, Hohai University, China; 
        Universidad de Buenos Aires Facultad de Ingenier\'\i a Instituto del Gas y del Petr\'oleo, Argentina; Department of Mathematics Purdue University Indiana. 
           \email{jesantos48@gmail.com}
         \and  
       J. M. Carcione \at 
          National Institute of Oceanography and Applied Geophysics - OGS, Trieste, Italy \\
      \email{jcarcione@inogs.it} 
       \and 
     G. Savioli \at 
    Universidad de Buenos Aires, Facultad de Ingenier\'\i a, IGPUBA,  Buenos Aires Argentina \\
  \email{gsavioli@fi.uba.ar}
    \and 
   P. Gauzellino \at Facultad de Ciencias Astron\'omicas y Geof\'\i sicas, Universidad Nacional de La Plata, Argentina \\
 \email{gauze@fcaglp.fcaglp.unlp.edu.ar}
  \and 
   A. Ravecca \at 
   Universidad Nacional Arturo Jauretche, Argentina.
   \email{amravecca@gmail.com}
  \and 
   A. Moras \at Universidad de Buenos Aires, Argentina. 
   \email{morasalfredo@hotmail.com} 
 }

\date{Received: date / Accepted: date}

\baselineskip 15pt

\maketitle 

\begin{abstract}
A pandemic caused by a new coronavirus has spread worldwide, affecting Argentina. 
We implement an SEIR model to analyze the disease evolution in Buenos Aires and neighbouring cities.
The model parameters are calibrated using the number of casualties officially reported. 
Since infinite solutions honour the data, we show different cases. In all of them the reproduction ratio $R_0$ decreases after early lockdown, but then raises, probably due to an increase in contagion in highly populated slums. Therefore it is mandatory to reverse this growing trend in $R_0$ by applying control strategies to avoid a high number of infectious and dead individuals. The model provides an effective procedure to estimate epidemic parameters (fatality rate, transmission probability, infection and incubation periods) and monitor control measures during the epidemic evolution.
\end{abstract}




\section{Introduction}
\label{}
In March 9, 2020, the start of the new coronavirus (COVID-19) epidemic in Argentina was officially reported by the Argentinian Ministry of Health. By today, June 24, 2020, the number of cases is still rising. The majority of mathematical models that replicate illnesses outbreaks splits the population into compartments, to analyze how much time will it take for one group to evolve into another \cite{Hethcote00,Brauer17}. The mathematical details of these models can be seen in  \cite{Hethcote00,Keeling08,Diekmann13}. 
Here, we use a Susceptible-Exposed-Infected-Removed (SEIR) model, consisting of  a system of  first-order ODE’s to describe the spread of the virus, compute the number
of infected individuals and  estimate the death toll. As examples, the SEIR model has been   successfully applied to study the transmission dynamics of tuberculosis \cite{Athithan13} and varicella \cite{Zha20}.    
It is important to clarify that the E class is infected but has not the symptoms of the disease, because they are incubating it. They will have symptoms when they pass to class I. Individuals in class I may not have symptoms (asymptomatic), but they are infectious, while those in class E are not. Moreover, individuals in class E can move to R without showing symptoms, but they are infectious when they are in class I. 

The SEIR model has been applied by Carcione et al. \cite{Carcione20} to simulate the epidemic in the Lombardy Region (Italy), with approximately 16500 casualties reported to date. The significant threat COVID-19 carries finds its meaning in the elevated number of infected health-care workers, such as 20 \% of the cases in Italy.

In order to analyze the epidemic's behaviour, the model is calibrated using the number of dead individuals to date, which we consider more reliable than the number of infectious individuals. The model parameters are: probability of transmission per contact, incubation and infectious periods and fatality rate. 

Based on China, USA and European Union’s experience, we are certain that combining rapid diagnosis with isolation measures has a substantial effect on the epidemic’s dynamics. Evaluating and quantifying the effectiveness of these methods is extremely important  \cite{Chowell03} and numerical simulators contribute to achieve this goal.  Dekhordi et al. \cite{Dehkordi20} presented a case study of  COVID-19 using an 
statistical analysis of  data from countries in Asia and Europe  like 
China, Italy and  Spain, among  others,  to characterize the dynamics of the pandemia.   

Summarizing, the numerical simulation aim is to provide an effective procedure to model 
the virus diffusion over time and to analyze the effectiveness of administrative measures. The ODE’S system solution is obtained employing a forward Euler scheme \cite{Carcione14} and we assume that natural deaths and births are balanced.
In this way  the peak of infected and dead individuals per day as a function of time can be predicted based on the parameters estimated during calibration. 

\section{The SEIR differential  model}
\label{}
This work  uses the SEIR epidemic model \cite{Hethcote00,Keeling08,Diekmann13,Zha20,Alsho04} to study the time evolution of the COVID-19 epidemic in Argentina.
The model considers a total (initial) population, $N_0$, composed of 
four compartments: susceptible, $S (t)$, exposed, $E (t)$, infected, $I(t)$ and recovered, $R (t)$, with 
$t$ being the time variable.
 
The initial value problem for the SEIR ODE's system is formulated as follows:
\begin{eqnarray}\label{eq1}
&&\dot S = \Lambda - \mu S - \beta S \dfrac{I}{N},\\ 
&&\dot E = \beta S \dfrac{I}{N} -  (\mu + \epsilon) E, \nonumber\\ 
&&\dot I = \epsilon E  - (\gamma + \mu + \alpha) I, \nonumber\ \\ 
&&\dot R = \gamma I - \mu R,  \nonumber\
\end{eqnarray}
with initial conditions $S(0), E(0), I(0)$ and $R(0)$.
In equation \eqref{eq1} the time derivative is denoted by a dot above the variable and $N$ is the number of live individuals at time $t$, e.g. $N = S + E + I + R \le N_0$.

The coefficients in  \eqref{eq1} are: the  birth rate $\Lambda$, the  natural per-capita death rate $\mu$, the virus induced average fatality rate $\alpha$ (its inverse is the  life expectancy of an individual in the infected class)  and the probability of disease transmission per contact $\beta$. Moreover,  $1/\gamma$  and $1/\epsilon$ are 
the infection and incubation periods, respectively. All of these coefficients have units  of (1/time).

The traditional SIR model \cite{Kumar20} is obtained selecting  $\Lambda = \mu =  0 $ and $\epsilon = \infty$ while if $\Lambda$ and $\mu$ are not zero, the model is termed endemic SIR \cite{Allen17}. However, as the SIR model has no exposed compartment, then it would not be proper using it for infections with $\epsilon$ values such as those of the COVID-19. 

Concerning the meaning of the variables in \eqref{eq1}, $S$ is the number of humans susceptible to be exposed and $E$ is the actual number of exposed individuals (individuals in which the disease is latent; they are infected but not infectious). Individuals move from $S$ to 
$E$ depending on the number of contacts with $I$ individuals, multiplied by the probability of infection ($\beta$). Furthermore,  exposed ($E$) become infected ($I$) with a rate $\epsilon$ and infected become recovered ($R$) with a rate $\gamma$. Since lifelong
immunity is assumed, people in the $R$ class do not move back to the $S$ class. 
Because of the relatively short period of the epidemic, it is assumed that  
$ \Lambda = \mu N$, so that the deaths balance
the newborns.

The deceased individuals $D(t)$ are computed as $D(t) = N_0 - N(t)$, therefore  
the dead people per unit time $\dot D (t)$, can be obtained as \cite{Sen17}:
\begin{equation} \label{eq2}
\dot D (t) = \alpha I (t). 
\end{equation} 
if the deaths balance
the newborns.

An important measure to quantify the virus expansion is the basic reproduction ratio, $R_0$, which estimates the average number of secondary cases caused by an already infected person. For the SEIR model, $R_0$ is given by  \cite{Zhang13}
\begin{equation} \label{31}
R_0 = \frac{\beta \epsilon}{(\epsilon+\mu) (\gamma+ \alpha + \mu)} 
\end{equation}

The basic reproduction ratio is used to estimate the virus spread, establishing $R_0$ = 1 as a stability limit: if $R_0 > 1$  the disease invade the population   while if $R_0 < 1$  the disease disappears.
 Al-Sheikh \cite{Alshe12}  analyzes in detail the behavior of the SEIR models in terms of  $R_0$. 

Another measure is the infection fatality rate (IFR), defined as  
\begin{equation} \label{IFR}
\mbox{IFR} \ (\%) = 100 \cdot \frac{D_\infty}{R_\infty + D_\infty}, 
\end{equation}
where $R_\infty + D_\infty$ represents the final number of infected individuals ($t \rightarrow \infty$ refers to the epidemics ending). 

Using the last equation (\ref{eq1}) (with $\mu$ = 0) and equation (\ref{eq2}), we obtain 
\begin{equation} \label{IFR1}
{\rm IFR} \ (\%) = 100 \cdot \frac{\alpha}{\alpha + \gamma} , 
\end{equation}
where this relation is always valid, not just at the epidemics ending.   

A third usually reported (time-dependent) coefficient  is the case fatality rate (CFR), such that CFR $>$ IFR, since this rate underestimates the number of infected individuals. 

Equations \eqref{eq1}  are discretized using a forward Euler discretization scheme with 
a time step  $\Delta t = 0.01$  day. The discrete  solution $(S^n, E^n, I^n, R^n)= (S, E, I ,R)(n \Delta t)$, 
$n\ge 1$ of this time discretization procedure yields positive and bounded solutions  \cite{Brauer17}. Furthermore, the solution  converges to an equilibrium, i.e., $S^n+R^n+D^n = S_\infty+R_\infty+D_\infty  = N_0$ for $t \rightarrow \infty$.  

A sensitivity analysis of the model to changes in its parameters is presented 
in \cite{Carcione20}, 
assuming that the parameters do not vary during the epidemic.
It is observed that higher values of the incubation period ($\epsilon^{-1}$) delay the epidemic,
and increasing the  infectious period ($\gamma^{-1}$) induces the same effect.
Furthermore, when more individuals are initially exposed ($E(0))$, the intensity of the peak does not change, but anticipates the epidemic.
Other results indicate that if $R_0$ does not change during the epidemic, the peak of infected people is  hardly sensitive to the initial number of infected individuals, and  an earlier lockdown highly reduces the number of dead individuals.

\section{The COVID-19 epidemic in the RMBA}
Next, we attempt  to model the COVID-19 epidemic in the RMBA, comprising  the city of Buenos Aires and neighbouring cities, with a population $N_0$ = 14839026 individuals.  
For this purpose, we use  as reliable data 
the total number of casualties from day 1 (March 9, 2020) to day 108 (June 24).  
The reported infected people cannot be used for calibration, since at present 
the number of asymptomatic, undiagnosed infectious individuals is unknown.
The number of death individuals is also uncertain, since there can be an under-ascertainment of deaths, but the error is much smaller than the error related to the infected individuals.
The number of dead individuals reported officially could possibly be underestimated due to undeclared cases. Thus, we also consider a case with 100 \% more dead people, compared to the official figures.  

Predictions of high accuracy are not possibly due to the lack of information about 
the probability of the disease transmission, characteristics of the disease and initial conditions 
of the SEIR model. We assume $\mu^{-1}$ = 3.6 $\times$ 10 $^{-5}$/ day, corresponding  to 
a life expectancy of  76 years.  Parameter  $\beta$ varies as a 
piecewise constant function in  time intervals [$t_0$, $t_1$], [$t_1$, $t_2$] $\ldots$ [$t_{L-1}$, $\infty$],
with changes associated with administrative measures taken by the state (such as lockdown) and behavior of the population. 
In this case, $t_0$ = 1 day,  
$t_1$ = 31 day and $t_2$ = 50 day, i.e., $L$ = 3, since after $t_1$ (April 8), home isolation, social distancing and partial Nation lockdown started to be effective, as indicated by an inflection point in the curve of casualties, and after $t_2$ (April 27), the situation became worst with an increase in the slope of the curve. 

The SEIR model parameters, $\alpha$, $\beta$, $\epsilon$, $\gamma$,  together with the initial exposed individuals $E(0)$ are estimated fitting the number of fatal victims. This constitutes an inverse problem with infinite solutions. A Quasi Newton approximation technique for nonlinear least squares problem with the formula of Broyden-Fletcher-Goldfarb-Shanno is applied  \cite{Gill81}. This technique was successfully applied to estimate parameters in reservoir engineering \cite{Savioli94}.
The L$^2$-norm of the differences between simulated and measured casualties is minimized and yields $E(0)$ and the model parameters on each time interval. In fact, while $\alpha$,   
$\epsilon$ and $\gamma$ remain constant throughout the entire simulation, 
$\beta$ varies in the intervals  [$t_0$, $t_1$],  [$t_1$, $t_2$] and  [$t_2$, $\infty$].

We show the results of five different cases that honour the data. The parameters constraints, initial values and results for the five cases are displayed in Table I.  All the cases assume an initial number of infectious individual $I(0)$ = 100, although this value may also affect the result.  
The last column do not correspond to variables but indicates the infected individuals at the end of the epidemic, i.e., $I_\infty = R_\infty + D_\infty$, the day of the last infected individual (the end of the epidemic in theory), and the death toll $D_\infty$.  
The results are very sensitive to variations of the parameter $\beta$, and consequently those of $R_0$, mostly due to the impact of the performed intervention strategies. The tragic situation predicted by the model is strongly influenced by the behaviour of $R_0$ during the latter period, after day 50. Therefore, a reduction of $R_0$ is essential to avoid this situation. 

Figure 1 and 2 show the fit and extended curves corresponding to Case 1, which 
predicts an initial $R_0$ = 3.34, decreasing to 0.95 in April 8, after the lockdown, and increasing to 1.59 after April 27, most probably due to an increase of the contagion in highly populated slums. This case, which is characterized by a long incubation period of approximately 11 days, predicts an IFR = 1.92 \% and a very high death toll (nearly 179 K) and 9.4 million infected individuals at the end of the epidemic. However, this is due to the last $R_0$ trend that can be inverted by implementing more isolation. In fact, if the value of $R_0$ = 0.95 had been maintained after April
27 only 1881 casualties and 92955 infected individuals would have occurred at
the end of the epidemic.
In the situation shown in Case 1, the maximum number of infected individuals is approximately 440000 
people at day 287 (December 21, 2020). 
If only 1 \% of these individuals require intensive care, this amounts to 4400 humans, a number that can overload the capacity of the hospitals.
The other cases honour the data with smaller incubation periods between 3.5 and 5 days, and predict less than 1/2 of casualties, compared to Case 1, with IFR between 0.5 \% and 1.15 \% approximately. 
For instance, Case 2 (Figure 3) which 
predicts an initial $R_0$ = 1.90, decreasing to  $R_0$ = 1.02 (approximately the stability limit) in April 8 , and increasing to 1.24 after April 27. Figure 3b shows a peak of 134 thousand infected individuals at 242 days (November 6, 2020), which implies 1340   
patients if 1 \% require intensive care, a more tractable amount. 
Figure 4 shows the results of Case 2 but keeping $R_0$ = 1.02 from day 31 (April 8). As can be seen, the epidemic would be under control with a minimum death toll (6500 individuals) and a minimum number of infected humans at the end of the epidemic (560 thousand people). Also the maximum number of infected individuals is very low, nearly 2200 individuals at day 386 (March 30, 2021). 

Since the reported number of deceased people could possibly be underestimated due to undeclared cases,  we also consider a case with 100 \% more casualties to date (Case 5 in Table I), giving  
IFR = 0.84 \% and values of the other parameters and infected and dead individuals similar to those of Cases 2, 3 and 4.

Data provided by literature can be compared with the values shown in Table I. The population's age affects the IFR and fatality rate. As an example an estimated IFR of  0.657 \% for the whole Chinese population grows to 3.28 for population over 60 yr age  (Verity et al. \cite{Verity20} -- Table I). If the amount of infected people is considerably superior than the reported cases, the mortality rate could be substantially lower than the official one, indicating COVID-19 is less lethal than SARS and MERS, despite of being far more contagious. Read et al. \cite{Read20} states an average value $R_0$ = 4, whereas Wu et al. \cite{Wu20}  get values within 1.8 and 2. As reported by Chowell et al. \cite{Chowell03}  IFR = 4.8 \% for SARS, and Verity et al. \cite{Verity20} state that CFR of SARS is superior than the one of COVID-19, with approximately 1.38 \%. Again, COVID-19 appears to be far more contagious. 

\vspace{0.2cm}

An extended approach consists in using time derivatives of fractional order to generalize the diffusion process. Such models include in a natural way both memory and non-local effects \cite{Mainardi10,Zeb14,Ahmed17,Chen20}. Indeed, replacing the first-order temporal derivative by a Caputo fractional derivative of non-natural order \cite{Caputo11} we obtain a new parameter to adjust the data: the derivative order. 
This modelling can be  performed by using  fractional derivatives computed with the
Gr\"unwald-Letnikov approximation, which is a generalization of the finite-difference derivative 
\cite{Carcione14,Carcione02} or solving the differential equations in the frequency domain \cite{Gauzellino14,Santos17}. 
Furthermore, the model can be made two-dimensional by including the spatial diffusion of the virus \cite{Naheed14,Qin14,Laaroussi19} to model local outbreaks and be able to isolate them. 
The approach can be based on a finite-element method in the space-frequency domain with domain decomposition. This numerical procedure has already been applied to wave propagation in 2D and 3D media in geophysics \cite{Santos17}. Alternatively, the Fourier pseudospectral method, to compute the spatial derivatives, combined with time-domain fractional derivatives, can be used to solve the 
space-time diffusion equation of the epidemic \cite{Carcione14,Carcione13}.
Moreover, there are more complex versions of the SEIR model as, for instance, including a quarantine class 
and a class of isolated (hospitalized) members \cite{Brauer17}. 
Finally, since signals propagate instantaneously in diffusion equations, the model predicts that there are more infectious humans (I class) than actual before the latent period and at late stages  of the epidemic. Solutions to this problem can be found, for instance, in Keeling and Rohani \cite{Keeling08} -- Section 3.3. 

 \section{Conclusions}
The SEIR epidemic model is implemented to simulate the time evolution of the COVID-19 epidemic in Argentina, specifically the ``Regi\'on Metropolitana de Buenos Aires" (RMBA), where the situation is more critical compared to other parts of the country. We calibrate the model parameters by using the number of  officially reported casualties, considered more reliable than the number of infected individuals. 
The simulation attempts to provide a simple but effective procedure to model 
the virus diffusion over time, in view of the lack of knowledge of many variables related to the 
epidemic. 

At present, the epidemic in the Buenos Aires area is under control due to the early lockdown, but we found that the reproduction ratio first decreased and then increased, causing a drastic prediction of the death toll if this trend persist in the future. In general, the incubation and infectious periods are in the range 4-5 days and 3-4 days, respectively, and the infection fatality rate (IFR) between 
0.5 \% and 2 \%. A case with an incubation period of 11 days yields approximately three times more casualties at the end of the epidemic, and the infected individuals will be between 4 million and 9 million people if the increasing $R_0$ trend is not inverted. 
 
We show how the effectiveness of the lockdown, the incubation and infectious periods, the probability of transmission and the initially exposed individuals affect the 
evolution of the epidemic. More complex models, i.e., with more classes or compartments and considering spatial diffusion, can be used in the future when some of the properties of the virus can be established more accurately, mainly the incubation and infectious periods and fatality rate.


\vspace{1cm}


{\bf Acknowledgements}: 
This work was partially funded by ANPCyT, Argentina (PICT 2015 1909) and Universidad de Buenos
Aires (UBACyT 20020160100088BA).
We are grateful to Nuria Sarochar, who helped us 
to evaluate the parameters of the SEIR model.

\newpage

\begin{center}
Table I. Constraints and initial--final values of the inversion algorithm.
\end{center}

\small
\[
\begin{tabular}{|c|c|c|c|c|c|c|c|c|c|}
\hline
  Case & Variable $\rightarrow$  
  & $\alpha$ 
  & $\beta_1$ 
  & $\beta_2$ 
  & $\beta_3$
  & $\epsilon^{-1}$ 
  & $\gamma^{-1}$ 
  & $E(0)$ 
  & $I_\infty$ (K)
  \\
 & 
 & day$^{-1}$ 
 & day$^{-1}$ 
 & day$^{-1}$  
 & day$^{-1}$   
& day 
& day  
& 
& $L$ (day)
\\
& 
 &  
 &   
 &   
&  
&
&   
& 
& $D_{\infty}$ (K) 
\\
\hline
\hline
 & Lower bound   
 & 10$^{-5}$ 
 & 10$^{-6}$ 
 & 10$^{-6}$ 
  & 10$^{-6}$  
 & 2 
 & 2 
 & 1 
 & \\
 & Upper bound   
 & 10$^{-1}$ 
 & 10$^{3}$
 & 10$^{3}$ 
 & 10$^{3}$  
 & 20
 & 20 
 & 10$^4$  
 & \\
 & Initial value  
 & 0.006 
 & 0.7 
 & 0.7 
 & 0.7
 & 10 
 & 10
 & 100 
 & 
 \\
\hline
 1.1
 & Final value  
 & 0.00285 
 & 0.5048 
 & 0.1434
 & 0.2400
 & 11.06
 & 6.74
 & 162
 & 9417
 \\
 1.2
 & IFR 
 & 1.92 \% 
 & 
 & 
 & 
 &
 &  
 & 
 & Sep 14 2021\\
 1.3
 & $R_0$ 
 & 
 & 3.34
 & 0.95
 & 1.59
 & 
 &  
 & 
 & 179
 \\
\hline
 2.1
 & Final value  
 & 0.00323 
 & 0.5799 
 & 0.3100
 & 0.3800
 & 4.29
 & 3.31
 & 44
 & 5382
 \\
 2.2
 & IFR 
 & 1.07 \% 
 & 
 & 
 & 
 &
 &  
 & 
 & Feb 3 2022\\
 2.3
 & $R_0$ 
 & 
 & 1.90
 & 1.02 
 & 1.24
 & 
 &  
 & 
 & 57 
 \\
 \hline
 3.1
 & Final value  
 & 0.00322 
 & 0.5899 
 & 0.2500
 & 0.3630
 & 4.44
 & 3.58
 & 10
 & 6011
 \\
 3.2
 & IFR 
 & 1.15 \% 
 & 
 & 
 & 
 &
 &  
 & 
 & Jan 1 2022 \\
 3.3
 & $R_0$ 
 & 
 & 2.09
 & 0.88 
 & 1.28
 &  
 &  
 & 
 & 69
 \\
 \hline
4.1
 & Final value 
 & 0.00150
 & 0.5000 
 & 0.3640
 & 0.3800
 & 3.45
 & 3.13
 & 490
 & 4344
 \\
 4.2
 & IFR 
 & 0.47\% 
 & 
 & 
 &
 & 
 &  
 & 
 & May 27 2021 \\
 4.3
 & $R_0$ 
 & 
 & 1.56
 & 1.13 
 & 1.18
 &  
 &  
 & 
 & 20 \\
\hline
 5.1
 & Final value$^{(*)}$ 
 & 0.0023 
 & 0.4400 
 & 0.3200
 & 0.3450
 & 4.65
 & 3.65
 & 500
 & 5439
 \\
 5.2
 & IFR 
 & 0.84 \% 
 & 
 & 
 & 
 &
 &  
 & 
 & March 12 2022 \\
 5.3
 & $R_0$ 
 & 
 & 1.59
 & 1.16 
 & 1.25
 &  
 &  
 & 
 & 46
 \\
\hline
\end{tabular}
\]

\small
\footnotesize

\baselineskip 10pt
\hspace{0.1cm} $I(0)$ = 100 

\hspace{0.1cm} ${(\ast)}$ Doubling the number of casualties. 

\hspace{0.1cm} The values of $\beta$ refer to the periods (in days): [1, 31], [31, 50] 
and [50, $\infty$] (in days). 

\hspace{0.1cm} $I_\infty$ indicates the total infected individuals at the end of the epidemic.

\hspace{0.1cm} $L$, the day of the last infected individual is obtained when $I < 1$.

\hspace{0.1cm} $D_\infty$ is the death toll at the end of the epidemic.

\hspace{0.1cm} Read et al. \cite{Read20} report the mean values $\epsilon^{-1}$ = 4 days and $\gamma^{-1}$ = 3.6 days.

\hspace{0.1cm} Lauer et al. \cite{Lauer20} report $\epsilon^{-1}$ = 5.1 days.

\hspace{0.1cm} Ferguson \cite{Ferguson20} estimate an average IFR = 0.9 \%.

\baselineskip 15pt

\newpage

\centerline{\bf List of Figure Captions}

\vskip1cm

Figures 1a,1b: RMBA case history. Dead individuals (a) and number of deaths per day (b), where
the symbols dots represent the data. The solid line corresponds to Case 1 in Table I.

\vskip1cm

Figures 2a,2b: Number of individuals (a) and deaths (b), corresponding to Case 1 in Table I. In (b) the
solid and dashed curves refer to the accumulated deaths and deaths per day. The peak of
infected individuals (and deaths per day) occurs at approximately day 287 (December 21, 2020).

\vskip1cm

Figures 3a,3b : Number of individuals (a) and deaths (b), corresponding to Case 2 in Table I. In (b) the
solid and dashed curves refer to the accumulated deaths and deaths per day. The peak of
infected individuals (and deaths per day) occurs at day 242 (November 6, 2020).

\vskip1cm

Figures 4a,4b: Number of individuals (a) and deaths (b), corresponding to Case 2 in Table I, but
maintaining the reproduction ratio $R_0$ = 1.02 after day 31, i.e., $\beta_3 = \beta_2$. 
In (b) the solid and dashed 
curves refer to the accumulated deaths and deaths per day. The peak of infected individuals
(and deaths per day) occurs at day  386 (March 30, 2021).

\newpage


\begin{figure}
\vskip1cm
\label{fig1}
\includegraphics[scale=0.35]{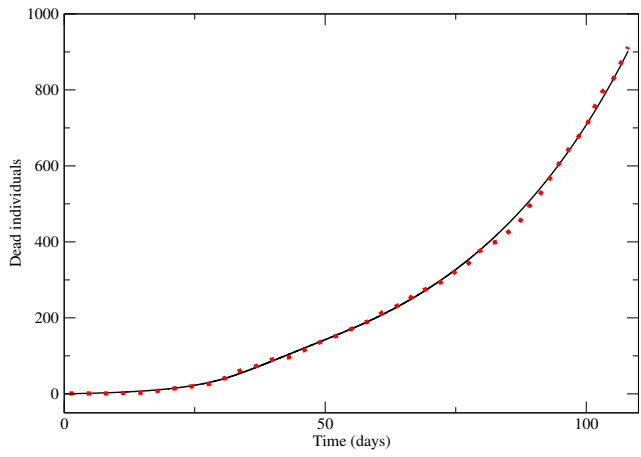}
\hskip-8.5 cm (a)\\ 
\vskip2cm
\includegraphics[scale=0.35]{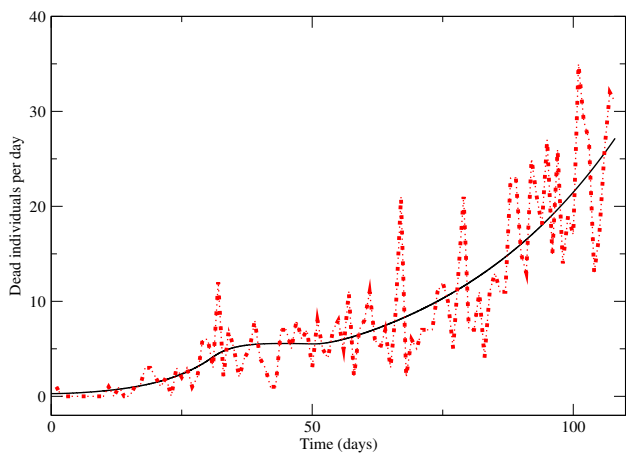} 
\hskip-8.5 cm (b)
\\
\vskip.2cm
\caption{RMBA case history. Dead individuals (a) and number of deaths per day (b), where
the symbols  dots represent the data. The solid line corresponds to Case 1 in Table I.}
\vspace{1cm}
\end{figure}

\begin{figure}
\vskip1cm
\label{fig2}
\includegraphics[scale=0.35]{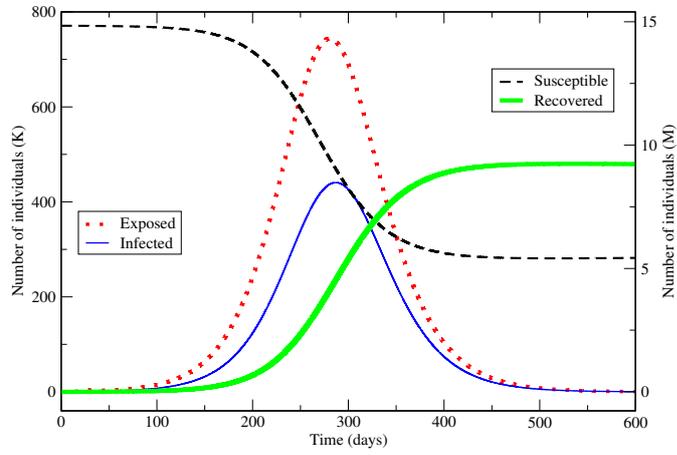}
\hskip-9.cm (a)\\ 
\vskip2cm
\includegraphics[scale=0.35]{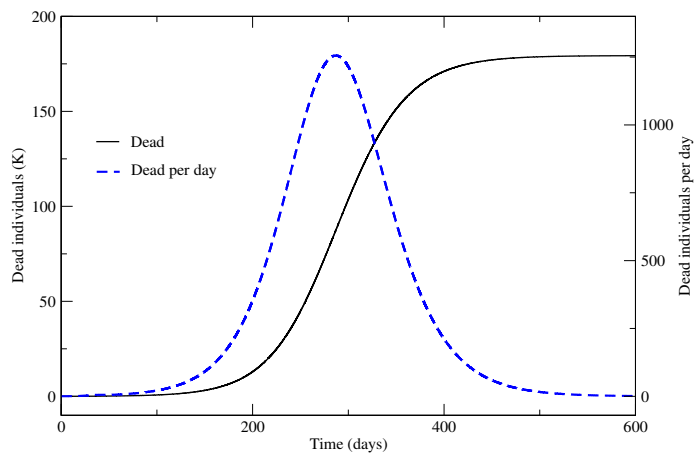} 
\hskip-9.cm (b)
\\
\vskip.2cm
\caption{Number of individuals (a) and deaths (b), corresponding to Case 1 in Table I. In (b) the
solid and dashed curves refer to the accumulated deaths and deaths per day. The peak of
infected individuals (and deaths per day) occurs at approximately day 287 (December 21, 2020).}
\vspace{1cm}
\end{figure}

\begin{figure}
\vskip1cm
\label{fig3}
\includegraphics[scale=0.35]{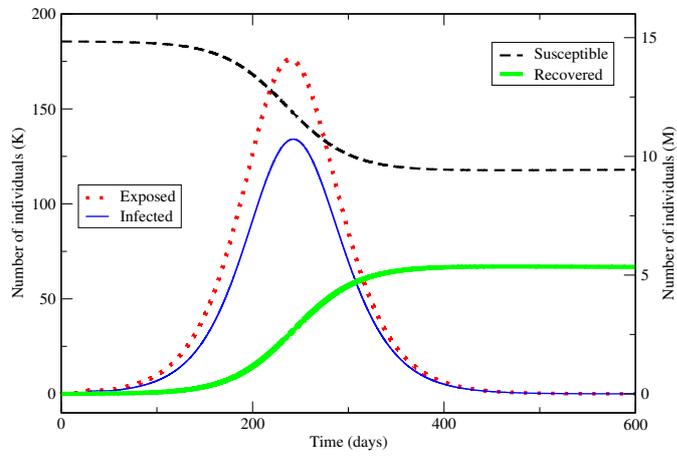}
\hskip-9.cm (a)\\ 
\vskip2cm
\includegraphics[scale=0.35]{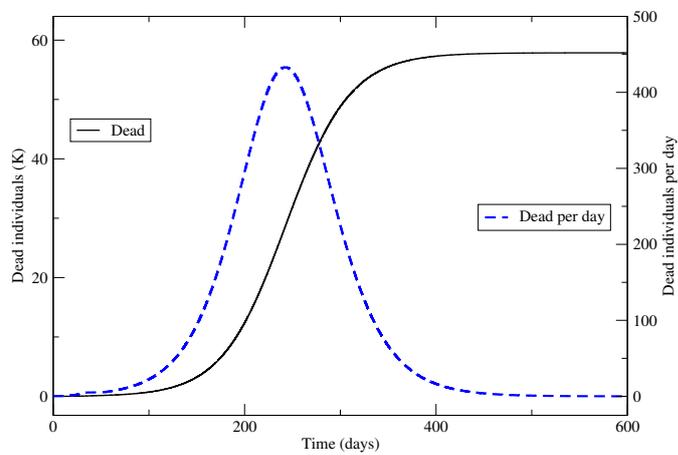} 
\hskip-9.cm (b)
\\
\vskip.2cm
\caption{Number of individuals (a) and deaths (b), corresponding to Case 2 in Table I. In (b) the
solid and dashed curves refer to the accumulated deaths and deaths per day. The peak of
infected individuals (and deaths per day) occurs at day 242 (November 6, 2020).}
\vspace{1cm}
\end{figure}

\begin{figure}
\vskip1cm
\label{fig4}
\includegraphics[scale=0.35]{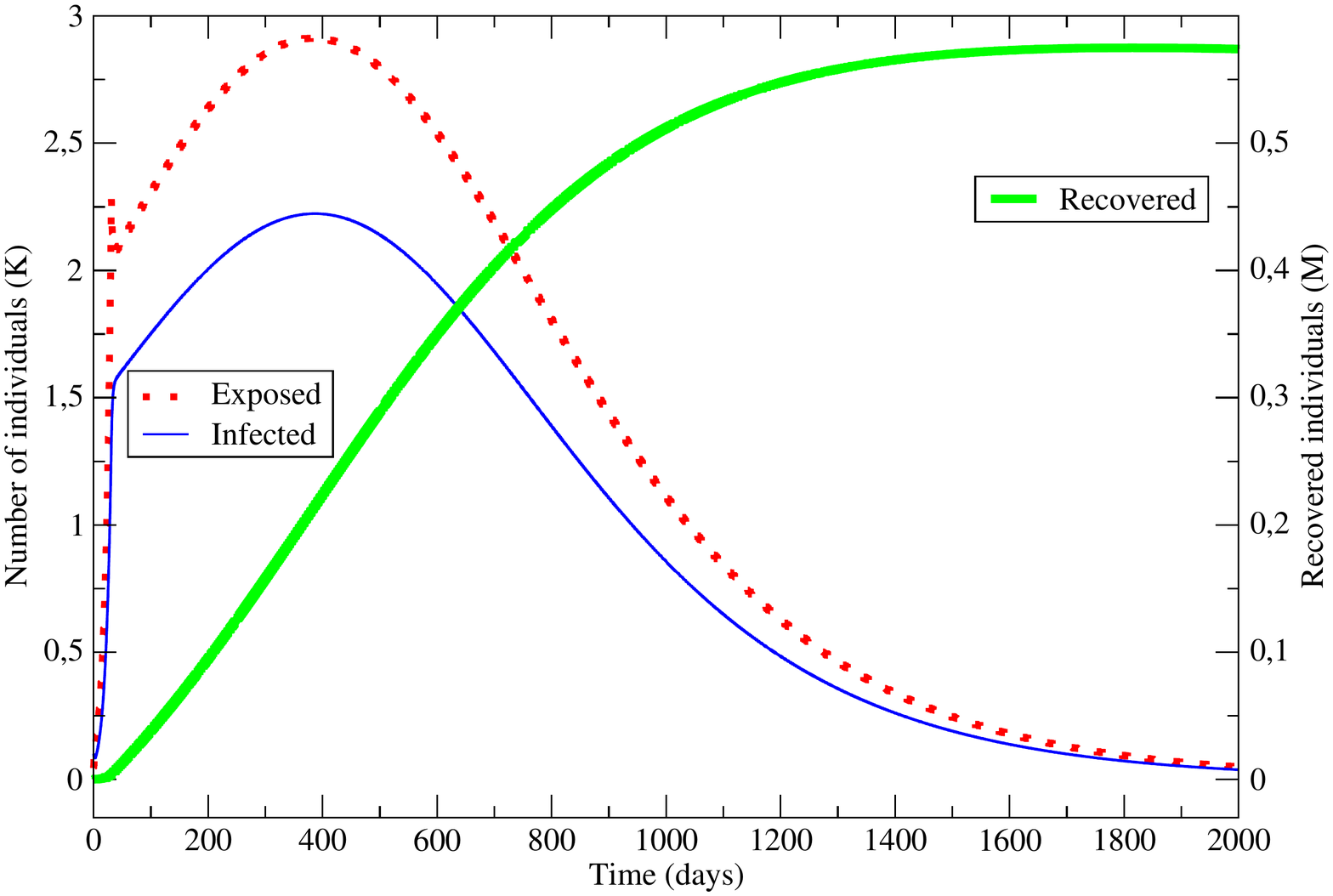}
\hskip-9.cm (a)\\ 
\vskip2cm
\includegraphics[scale=0.35]{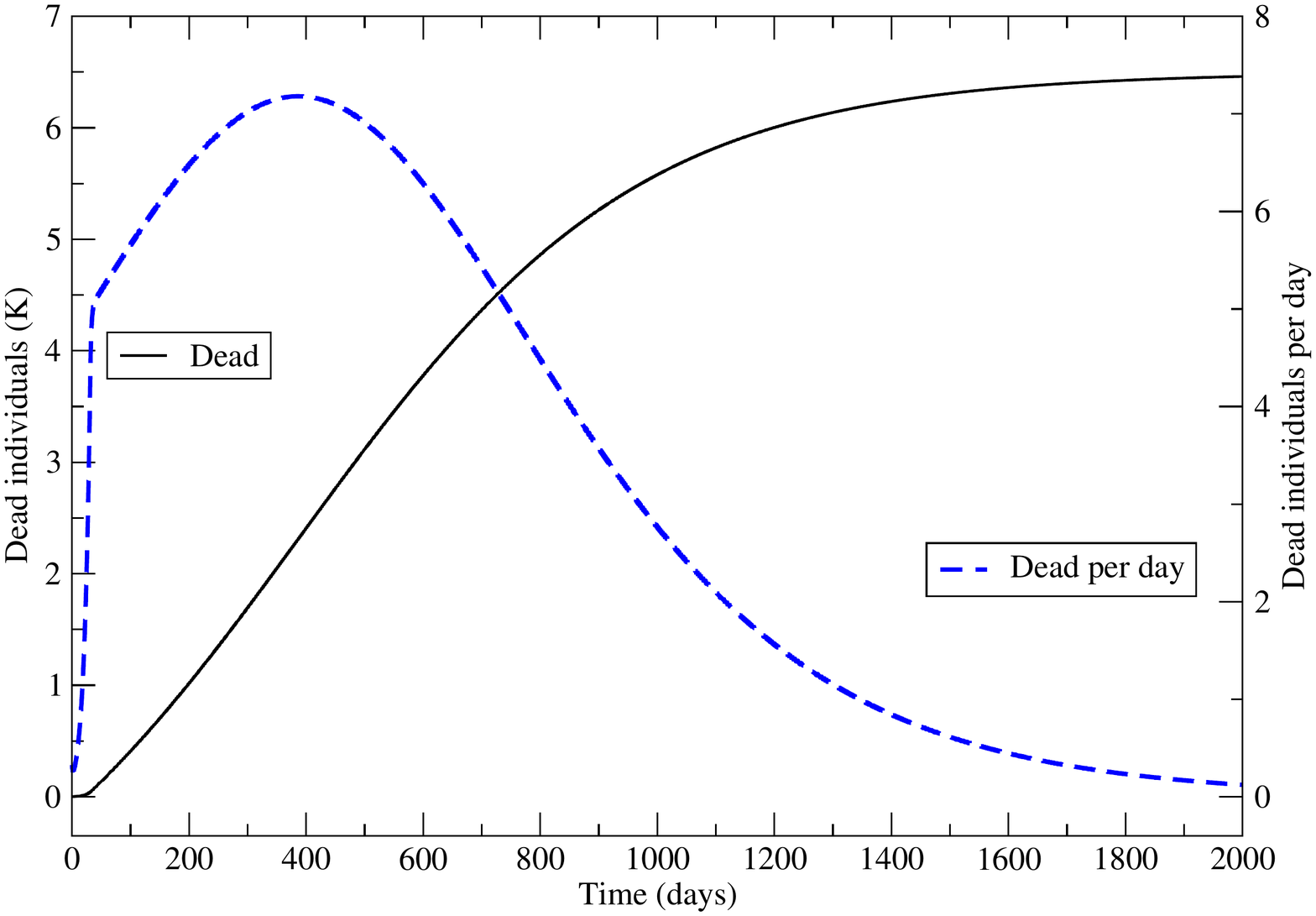} 
\hskip-9.cm (b)
\\
\vskip.2cm
\caption{Number of individuals (a) and deaths (b), corresponding to Case 2 in Table I, but
maintaining the reproduction ratio R 0 = 1.02 after day 31, i.e., $\beta_3 = \beta_2$. 
In (b) the solid and dashed 
curves refer to the accumulated deaths and deaths per day. The peak of infected individuals
(and deaths per day) occurs at day  386 (March 30, 2021).}
\vspace{1cm}

\end{figure}


\newpage
\vspace{5mm}


\begin{thebibliography}{}
\vspace{0.2cm} 

\bibitem{Hethcote00}
Hethcote H (2000) The mathematics of infectious diseases. SIAM Review 42: 599--653

\bibitem{Brauer17}
Brauer F (2017) Mathematical epidemiology: Past, present, and future. Infectious Disease Modelling 2(2): 113--127 
 
\bibitem{Keeling08}
Keeling M, Rohani P (2008) Modeling infectious diseases in humans and animals. Princeton University Press, Princeton 


\bibitem{Diekmann13}
Diekmann O, Heesterbeek H, Britton T (2013) Mathematical tools for understanding infectious disease dynamics. 
Princeton Series in Theoretical and Computational Biology, Princeton University Press, Princeton 

\bibitem{Athithan13}
Athithan S, Ghosh M (2013) Mathematical modelling of TB with the effects of case detection and treatment. Int J Dynam Control 1: 223--230


\bibitem{Zha20}
Zha W, Pang F, Zhou N, Wu B, Liu Y, Du Y, Hong X, Lv Y (2020) Research about the optimal strategies for prevention and control of varicella outbreak in a school in a central city of China: Based on an SEIR dynamic model. Epidemiology and Infection 148, E56,  \url{https://doi.org/10.1017/S0950268819002188}


\bibitem{Carcione20}
Carcione J, Santos J, Bagaini C, Ba J (2020)  
A simulation of a COVID-19 epidemic based on a
deterministic SEIR model.
Frontiers in Public Health , doi: 10.3389/fpibh.2020.00230, \url{https://arxiv.org/abs/2004.03575}


\bibitem{Chowell03}
Chowell G, Fenimore P, Castillo-Garsow M, Castillo-Chavez C (2003) SARS outbreak in Ontario, Hong Kong and Singapore: the role of diagnosis and
isolation as a control mechanism. J Theor Biol 224: 1--8

\bibitem{Dehkordi20}
Dehkordi A, Alizadeh M, Derakhshan P, Babazadeh P, Jahandideh A (2020) Understanding epidemic data and statistics: A case study
of COVID‐19. J Med Virol ; in press, doi: 10.1002/jmv.25827.

\bibitem{Carcione14}
Carcione J (2014) Wave Fields in Real Media. Theory and numerical simulation 
of wave propagation  in anisotropic, anelastic, porous and electromagnetic media,
3rd edition. Elsevier, Amsterdam


\bibitem{Alsho04}
Al-Showaikh F, Twizell E (2004) One-dimensional measles dynamics. Appl Math Comput 152: 169--194


\bibitem{Kumar20}
Kumar A, Goel K, Nilam (2020)  A deterministic time-delayed SIR epidemic model: mathematical modeling and analysis. Theory Biosci 139: 67--76. 

 \bibitem{Allen17}
Allen L (2017) A primer on stochastic epidemic models: Formulation,
numerical simulation, and analysis. 
Infectious Disease Modelling 2(2): 128--142  
 
 
\bibitem{Sen17}
De la Sen M, Ibeas A,  Alonso-Quesada S, Nistal R (2017) 
On a new epidemic model with asymptomatic and
dead-infective subpopulations with feedback controls
useful for Ebola disease. Discrete Dynamics in Nature and Society  
\url{https://doi.org/10.1155/2017/4232971}

\bibitem{Zhang13}
Zhang L, Li Y, Ren Q, Huo Z (2013) 
Global dynamics of an SEIRS epidemic model with constant
immigration and immunity. 
WSEAS Transactions on Mathematics 12: 630--640 

 
\bibitem{Alshe12}
Al-Sheikh S (2012) Modeling and analysis of an SEIR epidemic model with a
limited resource for treatment. Global Journal of Science Frontier Research, 
Mathematics and Decision Sciences 12(14): 57--66  



\bibitem{Gill81}
Gill P, Murray W, Wright M (1981) Practical Optimization. Academic Press, London
   
   
\bibitem{Savioli94}
Savioli GB, Bidner MS (1994) Comparison of optimization techniques for automatic history matching. Journal of Petroleum Science and Engineering 12(1): 25--35    
   
\bibitem{Verity20}
Verity R et al. (2020) 
Estimates of the severity of coronavirus disease 2019:
a model-based analysis. Elsevier Public Health Emergency Collection  
\url{https://doi.org/10.1016/S1473-3099(20)30243-7}


\bibitem{Read20}
Read J, Bridgen J, Cummings D, Ho A, Jewell C (2020) 
Novel coronavirus 2019-nCoV: early estimation of epidemiological parameters and epidemic predictions \url{https://doi.org/10.1101/2020.01.23.20018549}. 


\bibitem{Wu20}
Wu J, Leung K, Bushman M, Kishore N, Niehus N, 
de Salazar P,  Cowling B,  Lipsitch M, Leung G (2020)  Estimating clinical severity of COVID-19 from the transmission dynamics in Wuhan, China. Nature Medicine Letters  \url{https://doi.org/10.1038/s41591-020-0822-7}


\bibitem{Mainardi10}
Mainardi F (2010) Fractional Calculus and Waves in Linear Viscoelasticity. Imperial College Press, London


\bibitem{Zeb14}
Zeb A, Khan M, Zaman G, Momani S,  Erturk V (2014)  
Comparison of Numerical Methods of the SEIR Epidemic Model of Fractional
Order. Z Naturforsch 69a: 81 – 89,  DOI: 10.5560/ZNA.2013-0073


\bibitem{Ahmed17}
Ahmed E,  El-Saka H (2017) 
On a fractional order study of middle East respiratory syndrome corona virus (MERS-COV).
J Fractional Calculus  Appl 8(1): 118-126  


\bibitem{Chen20}
Chen Y, Cheng J, Jiang X,  Xu X  (2020) 
The reconstruction and prediction algorithm of the fractional
TDD for the local outbreak of COVID-19.  
\url{arXiv:2002.10302v1 [physics.soc-ph]}.



\bibitem{Caputo11}
Caputo M, Carcione J,  Cavallini F (2011) Wave simulation in biological
media based on the Kelvin-Voigt fractional-derivative stress-strain relation. Ultrasound
in Med \& Biol  37: 996--1004



\bibitem{Carcione02}
Carcione J, Cavallini F, Mainardi F,  Hanyga A  (2002) Time-domain seismic
modeling of constant $Q$-wave propagation using fractional derivatives. Pure and Applied
Geophysics 159(7): 1719--1736


\bibitem{Gauzellino14}
Gauzellino P, Carcione J, Santos J, Picotti S (2014)  A rheological
equation for anisotropic-anelastic media and simulation of synthetic seismograms. Wave
Motion 51: 743--757
     

     
\bibitem{Santos17}
Santos J, Gauzellino P (2017) Numerical simulation in applied geophysics. Lecture Notes in Geosystems Mathematics and Computing. Birkhauser, Cham


\bibitem{Naheed14}
Naheed A, Singh M, Lucy D (2014)  Numerical study of SARS epidemic model
with the inclusion of diffusion in the system. Applied Mathematics and Computation 229: 480--498


\bibitem{Qin14}
Qin W, Wang L, Ding X (2014)
A non-standard finite difference method for a hepatitis B virus
infection model with spatial diffusion. Journal of Difference Equations and
Applications DOI: 10.1080/10236198.2014.968565
     
\bibitem{Laaroussi19}
Laaroussi A, Ghazzali R, Rachik M, Benrhila S (2019) Modeling the spatiotemporal transmission of Ebola disease and optimal control: a regional approach. Int J Dynam Control 7: 1110--1124
     
\bibitem{Carcione13}
Carcione J, Sanchez-Sesma F, Luz\'on F, Perez Gavil\'an J (2013)  Theory
and simulation of time-fractional fluid diffusion in porous media. J. Phys. A: Math
Theor  46: 345501, 
doi:10.1088/1751-8113/46/34/345501

\bibitem{Lauer20}
Lauer S, Grantz K, Bi Q, Jones F, Zheng Q, Meredith H, Azman A, Reich N, Lessler J (2020) The incubation period of coronavirus disease 2019 (COVID-19) from
publicly reported confirmed cases: Estimation and application. Annals of Internal Medicine  
DOI: 10.7326/M20-0504




\bibitem{Ferguson20}
Ferguson N. et al. (2020) Impact of non-pharmaceutical interventions (NPIs) to reduce COVID-19 mortality and healthcare demand. \url{https://doi.org/10.25561/77482}

               

\end{thebibliography}
\end{document}